\documentclass[3p,times,twocolumn]{elsarticle}
 \biboptions{comma,sort&compress}
 
\usepackage{mathtools} 
\usepackage{amssymb}
\usepackage{graphicx}
\usepackage{float}
\usepackage[utf8]{inputenc} 
\usepackage{color}
\usepackage{wrapfig}
\usepackage{multirow} 
\usepackage{enumitem} 
\usepackage{amsmath}

\usepackage{fancyhdr} 
\usepackage{lipsum}
\usepackage{natbib} 
\usepackage[makeroom]{cancel} 
\newcommand{\lagd}{\mathcal{L}} 

\usepackage{slashed} 
\usepackage[table,xcdraw]{xcolor}
\usepackage[hidelinks]{hyperref}
\usepackage{physics}
\usepackage{makecell} 
\usepackage{adjustbox}
\usepackage{ecrc}

\usepackage{color}

\volume{00}

\firstpage{1}

\journalname{Nuclear and Particle Physics Proceedings}
\runauth{Stephan Narison}

\jid{nppp}
\jnltitlelogo{Nuclear and Particle Physics Proceedings}
\usepackage[figuresright]{rotating}

\begin{document}

\begin{frontmatter}

\title{Software tools for computing EW chiral amplitudes$^*$} 

\cortext[cor0]{Mini-Review talk presented at QCD22 - 25th International Conference in Quantum Chromodynamics (4-7/07/2022, Montpellier - FR).}

 \author[label1]{Javier Mart\'\i nez-Mart\'\i n
 \corref{cor1} 
 }
   \address[label1]{Departamento de F\'\i sica Te\'orica and IPARCOS, Facultad de Ciencias F\'\i sicas, 
   \\ 
   Universidad Complutense de Madrid, Plaza de Ciencias 1, Madrid 28040, España.
}
 \cortext[cor1]{Speaker. Talk based on Javier Mart\'\i nez-Mart\'\i n's Master Thesis~\cite{tfm}. Special thanks to my supervisor Juan Jos\'e Sanz-Cillero.  
This research is partly supported
by the Ministerio de Ciencia e Inovaci\'on under research grant PID2019-108655GB-I00/AEI/10.13039/501100011033; by the EU STRONG-2020 project under the program H2020-INFRAIA-2018-1 [grantagreement no. 824093], and IPARCOS grant "Ayudas de Máster IPARCOS-UCM/2021".
}

\ead{javmar21@ucm.es}

\author[label1]{Juan José Sanz-Cillero 
 }

\ead{jjsanzcillero@ucm.es}

\pagestyle{myheadings}
\markright{ }
\begin{abstract}
\noindent
We present an implementation of the Electroweak Chiral Lagrangian (also denoted as Higgs Effective Theory) in several high energy physics  \textsc{Mathematica} packages. In particular, we implement the bosonic part of the electoweak Lagrangian up to next-to-leading order for \textsc{FeynRules}, \textsc{FeynArts} and \textsc{FeynCalc}. These tools are publicly available and can be used to readily calculate the Feynman rules and amplitudes in the theory.   
 
\end{abstract}
\end{frontmatter}
\section{Introduction}
The Electroweak Effective Theory (EWET), also known as Higgs Effective Field Theory (HEFT) or Electroweak Chiral Lagrangian (EWChL)~\cite{LHCHiggsCrossSectionWorkingGroup:2016ypw}, is a non-linear generalization of the Standard Model Effective Field Theory (SMEFT)~\cite{Grzadkowski:2010es}. The formal aspects of the theory at leading (LO) and next-to-leading order (NLO) have been discussed previously in the literature~\cite{Kraus19,Pich17,Long80,Long81,Buch14,Fer93,Her94,Alonso:2012px}. For this review, we will be using the notation as found in Refs.~\cite{Kraus19,Pich17}. 

Here we present a series of packages~\cite{tfm} that automatizes the calculation of the vertices and interactions of the theory and can be easily modified to get terms of the LO and NLO Lagrangians with as many Goldstone fields as required. We have implemented this code using some existing packages of \textsc{Mathematica}:  \textsc{FeynRules} \cite{Rules}, \textsc{FeynArts} \cite{Arts} and \textsc{FeynCalc} \cite{Calc}. Some alternative implementations of the Lagrangian can be found in the \textsc{FeynRules} database \cite{RulesData}, but they are not meant to be modified nor extended with ease.

Our implementation of the theory is still a work in progress. Up to the present proceedings we have implemented the boson sector of the theory at LO and NLO in the chiral expansion~\cite{Kraus19,Pich17} (leaving out for now the fermion sector, color sector and other possible additions). For convenience, in~\cite{tfm}, we also provide a file with the vertices up to 6 particles for the effective Lagrangian up to NLO and, a smaller one, with the vertices up to 4 particles.

\section{The underlaying theory: The Chiral Lagrangian}

The Electroweak Chiral Lagrangian is based on the electroweak chiral symmetry breaking (EWSB) pattern of the scalar sector of the SM, $\mathcal{G}\equiv SU(2)_L \otimes SU(2)_R\rightarrow \mathcal{H}\equiv SU(2)_{L+R}$. For this effective theory we have used the particle content of the SM: the Higgs boson ($h$), the three electroweak (EW) Goldstone bosons ($\pi^a$) and the four EW gauge bosons ($W^{\pm}$,$Z$,$A$). In addition, the Higgs boson is incorporated as a scalar singlet with $m_h=125$ GeV (not part of a complex doublet together with the three Goldstones).

The following construction is summarized from the Master Thesis found as Manual in Ref.~\cite{tfm}, based on Refs.~\cite{Kraus19,Pich17}.

The Lagrangian is organized as as a low-energy expansion in powers of generalized momenta (derivatives and particle masses): 
\begin{equation}
    \lagd_{\text{EWET}}=\sum_{\hat{d}\geq 2}\lagd^{(\hat{d})} \, , \quad \lagd^{(\hat{d})}=\mathcal{O}(p^{\hat{d}})\, ,
\end{equation}
where the chiral dimension $\hat{d}$ reflects the infrared behaviour at low momenta (notice that the operators are not ordered according to their canonical dimension). The scale characterizing the EWSB will be defined as the vacuum expectation value ($v=(\sqrt{2}G_F)^{-1/2}=246$ GeV. 

This counting classifies the terms of the Lagrangian depending on how they contribute to the dimension of the amplitude $\mathcal{M}$ of a process. The order $p^{\hat{d}}$ of a diagram is given by the master formula 
\begin{equation}
   \mathcal{M}\sim p^{2(L+1)+\sum\limits_{\hat{d}} N_{\hat{d}}(\hat{d}-2)}\,,
\end{equation}
where $L$ is the number of loops and $N_{\hat{d}}$ the number of vertices of dimension $\hat{d}$. The $\mathcal{O}{(p^2)}$ diagrams provide the LO amplitude. They must contain only dimension $p^2$ vertices (from the LO Lagrangian $\lagd^{(2)}$) and no loops. We will also consider NLO diagrams, $\mathcal{O}{(p^4)}$. The diagrams that contribute at NLO must have one loop with only vertices of dimension 2 or no loop with exactly one vertex of dimension 4 (from the NLO term of the Lagrangian $\lagd^{(4)}$) and any number of vertices of dimension $p^2$.

The chiral dimension of the objects used to build our theory is as follows~\cite{Pich17}:
\begin{align}
        \frac{h}{v}, \, \frac{\pi^a}{v}, \, \frac{W^{a,\mu}}{v}, \, \frac{B^\mu}{v} &\sim \mathcal{O}(p^0) \, ,\\
    \partial_\mu,\, D_\mu,\, m_h, \, m_W, \, m_Z, \, g_W, \, g_1 &\sim \mathcal{O}(p) \, ,
\end{align}
where the gauge coupling constants $g_W$ and $g_1$ are also known in the literature as $g$ and $g'$ respectively. With the dimension of these objects, the reader can obtain the dimension of any other operator built with them.

In order to make the code easier to read, we have split the terms of the expansion into two parts: one denoted as $\lagd^{(\hat{d})}_{\text{Scalar}}$ including only the operators $u_\mu$, $h$ and the custodial breaking $\mathcal{T}$; and one $\lagd^{(\hat{d})}_{\text{FS}}$ including the field strength operators along with the previous ones. 
We have then,
\begin{equation}
   \hspace*{-0.5cm}  
   \lagd^{(\hat{d})}=\lagd^{(\hat{d})}_{\text{Scalar}}+\lagd^{(\hat{d})}_{\text{FS}}\,( \, + \, \text{fermionic / color terms}) \, .
\end{equation}

Let us now have a closer look at the fields appearing in the Lagrangian. Firstly, we have the Higgs, \textit{h}, which is a singlet under the $\mathcal{G}$ group transformations. 

Secondly, we build a $2\times 2$ Matrix containing the Goldstones fields $\pi^a$:
\begin{equation} \label{eq: U}
    U(\pi) = u(\pi)^2 = \exp{i\sigma^a\pi^a/v} \,.
\end{equation}

Then, for the gauge fields we defined the $2\times 2$ matrices,
\begin{equation}
    \hat{W}^\mu=-g_W\frac{\sigma^a}{2}W^{a,\mu}, \quad \hat{B}^\mu=-g_1\frac{\sigma^3}{2}B^{\mu} \, ,
\end{equation}
the covariant derivative,
\begin{equation}
    D_\mu U = \partial_\mu U - i \hat{W}_\mu U + i U \hat{B}_\mu \quad \, ,
\end{equation}
and the corresponding field-strength tensors
\begin{align}
    \hat{W}_{\mu\nu}&=\partial_\mu\hat{W}_{\nu}-\partial_\nu\hat{W}_{\mu}-i[\hat{W}_{\mu},\hat{W}_{\nu}]\, ,\\
    \hat{B}_{\mu\nu}&=\partial_\mu\hat{B}_{\nu}-\partial_\nu\hat{B}_{\mu}-i[\hat{B}_{\mu},\hat{B}_{\nu}]\, .
\end{align}
Some extra operators used here can be found in~\cite{tfm}.

Now, we build the EWChL. First, the FS leading order (LO) Lagrangian corresponds to the Yang-Mills (YM) Lagrangian for the EW gauge bosons:
\begin{equation}
    \lagd^{(2)}_{\text{FS}}=-\frac{1}{2g_W^2}\left\langle \hat{W}_{\mu\nu} \hat{W}^{\mu\nu}\right\rangle-\frac{1}{2g_1^2}\left\langle \hat{B}_{\mu\nu} \, , \hat{B}^{\mu\nu}\right\rangle \, ,
\end{equation}
where $\left\langle\cdot\right\rangle$ denotes the trace over the $SU(2)$ space.

Then, the bosonic Lagrangian:
\begin{align}
    \lagd^{(2)}_{\text{Scalar}}=&\frac{1}{2}\partial_\mu h \partial^\mu h - \frac{1}{2}m_h^2h^2-V\left({h}/{v}\right)+\\
    &+\frac{v^2}{4}\mathcal{F}_u\left({h}/{v}\right)\left\langle u_\mu u^\mu\right\rangle \, ,
\end{align}
with
\begin{align}
    V\left({h}/{v}\right)&=\frac{1}{2}m_h^2v^2\left[\sum_{n\geq 3}b_n\left(\frac{h}{v}\right)^n\right] \, ,\\
   \mathcal{F}_u\left({h}/{v}\right)&=1+\sum_{n=1}c_n^{(u)}\left(\frac{h}{v}\right)^n \, .
\end{align}
where we will also use the notation  $c_1^{(u)}\equiv2a$, $c_2^{(u)}\equiv b$.

At NLO, the $\mathcal{O}(p^4)$ operators must be considered alongside the one-loop corrections of the LO Lagrangian (they are of the same order). The NLO {\it FS}-type Lagrangian can be written as
\begin{align}
    \lagd^{(4)}_{\text{FS}}=& 
   \sum_{i=1}^{3}
    \left[\mathcal{F}_i\left({h}/{v}\right)\mathcal{O}_i+\tilde{\mathcal{F}}_i\left({h}/{v}\right)\tilde{\mathcal{O}}_i    \right] \nonumber \\
    &+\mathcal{F}_9\left({h}/{v}\right)\mathcal{O}_9+\mathcal{F}_{11}\left({h}/{v}\right)\mathcal{O}_{11} 
\, ,
\end{align}
and the NLO {\it scalar}-type Lagrangian as
\begin{align}
    \lagd^{(4)}_{\text{Scalar}}=&\sum_{i=4}^{8}\mathcal{F}_i\left({h}/{v}\right)\mathcal{O}_i+\mathcal{F}_{10}\left({h}/{v}\right)\mathcal{O}_{10} \, ,
\end{align}
where the coefficients $\mathcal{F}_i\left({h}/{v}\right)$ and $\tilde{\mathcal{F}}_i\left({h}/{v}\right)$ are analytical series of $h/v$ of the form 
\begin{align}
    \mathcal{F}_i\left({h}/{v}\right)=\sum_{n=0}^\infty\mathcal{F}_{i,n}\left(\frac{h}{v}\right)^n \, , 
\end{align}
with an analogous definition for $\tilde{\mathcal{F}}_i\left({h}/{v}\right)$. The operator $\mathcal{O}_i$ used to build these terms are collected in Ref.~\cite{tfm}.

\section{The software implementation}
\begin{figure}[]
\begin{center}
\includegraphics[width=0.38\textwidth]{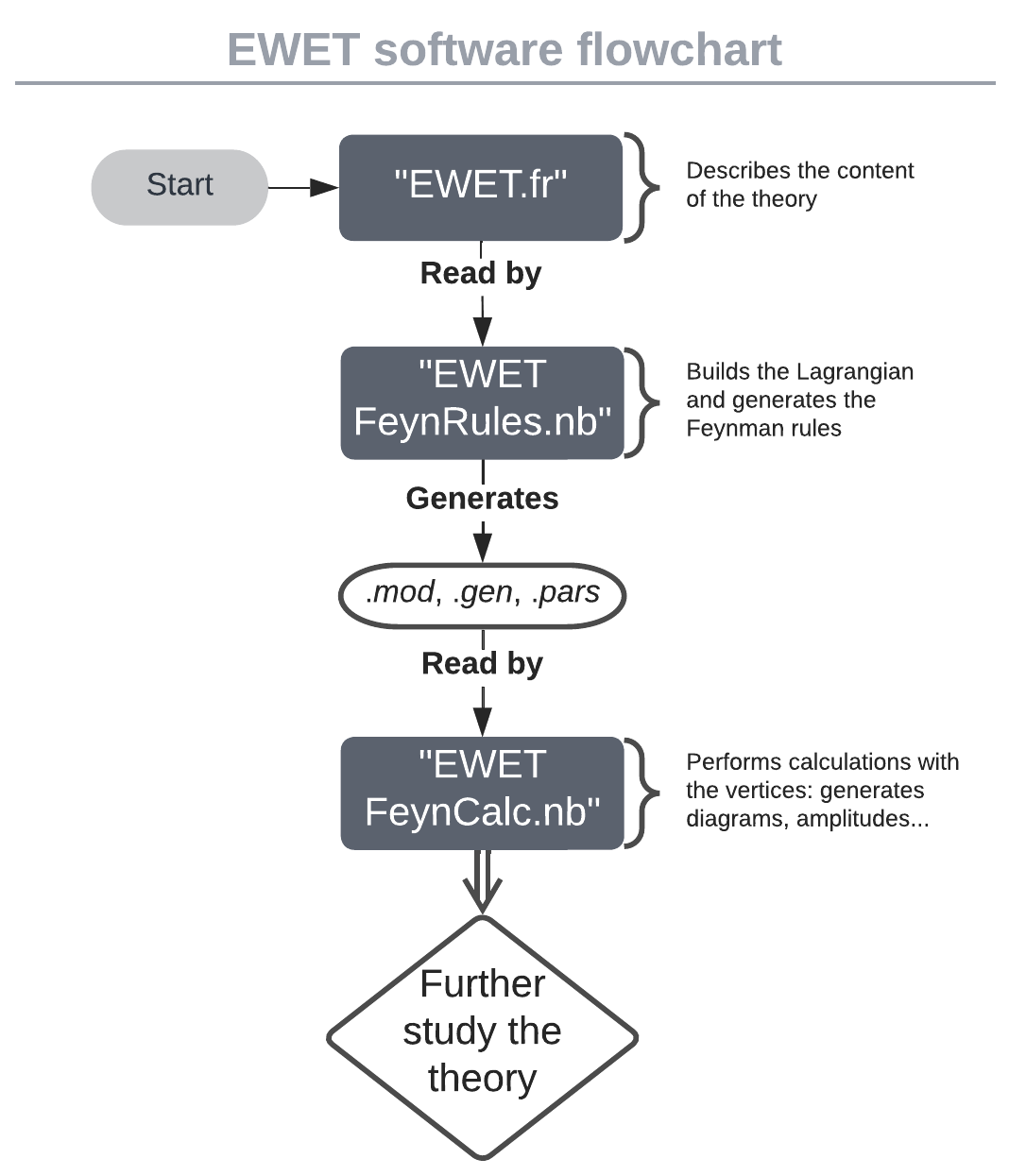}
\end{center}
\caption{Overview of the effective theory implementation} \label{fig: flow}
\end{figure}

The implementation has been made following the flowchart in Fig.~\ref{fig: flow}. In this review, we will cover the key aspects of our code, leaving further details for Ref.~\cite{tfm}. The files we talk about in this paper are available in GitHub \cite{tfm}.

First, we have built the model file of the theory in ``EWET.fr''~\cite{tfm}. This file contains the fields used to build the different operators and the parameters of the Lagrangian. We need to remark that the declared names are intuitively related to the quantity they represent. For example, the coupling $c_3^{(u)}$ is declared as \texttt{c3u} and the coupling $\mathcal{F}_{i,n}$ with $i=1$ and $n=2$ as \texttt{F1n2} (the $\tilde{\mathcal{F}}_{i,n}$ are represented by double \texttt{F} at the beginning of its name, for example \texttt{FF1n2} stands for $\tilde{\mathcal{F}}_{1,2}$). A definition of any particular parameter can be found within the model file.

Due to the complexity of the Lagrangian, we have opted to write a \textsc{Mathematica} notebook ``EWET FeynRules.nb''~\cite{tfm} where we describe the operators and the Lagrangian and get the Feynman rules of the theory. We have prepared the program to get every operator up to 6 fields, so that we get every vertex needed for studying LO (with tree level $\mathcal{L}^{(2)}$) and NLO (with 1 loop $\mathcal{L}^{(2)}$ and tree level $\mathcal{L}^{(4)}$) diagrams with up to 4 external legs. To avoid computing terms with a higher number of fields, we have declared a variable \texttt{epsf} which is attached to every field so that every term we compute is expanded up to \texttt{epsf}$^6$. 
Here the reader could extend the theory and get the operators up to any number of fields needed.

The computation of the vertices takes several minutes and then, returns 384 different vertices up to 6 particles for the Lagrangian up to NLO. This information is stored in the \textit{.mod}, \textit{.gen} and \textit{.pars} files under the name ``SixPartVertsLagr''\cite{tfm}.
In addition, the reader can find some simplifications of the theory vertices exported in the GitHub\cite{tfm} as explained in Ref.~\cite{tfm}.

Once we have the files \textit{.mod}, \textit{.gen} and \textit{.pars} generated thanks to the ``EWET FeynRules.nb''\cite{tfm} notebook, we do not need to execute that notebook again unless we need to change something from the theory (add fermions, $\mathcal{O}(p^6)$ operators...). The reader needs to take into account that executing that notebooks takes some computation time. Hence, using the provided \textit{.mod}, \textit{.gen} and \textit{.pars} is the fastest way to study the effective theory discussed in this paper.

Given that \textsc{Mathematica} needs to kill the kernel before executing a new package and that we do not want to keep using the prior notebook, we have created for illustration a new notebook ``EWET FeynCalc.nb''\cite{tfm} which runs \textsc{FeynCalc} (with a \textsc{FeynArts} add-on). With this notebook we can draw the diagrams for a given process, get the amplitudes and simplify them just by specifying the location of the model \textit{.mod} and the generic model \textit{.gen}.

The option to set up the Mandelstam variables in \textsc{FeynCalc} needs to be carefully selected for every diagram. When calculating the different amplitudes, we have considered the different definition of the Mandelstam sum considering the particles mass (note that Goldstones are massless in our selected Landau gauge).

With the implementation finished, we have checked its validity comparing some of our vertices and amplitudes (both at LO and NLO) with others found in the literature~\cite{Rafa,Clau} (see Ref.~\cite{tfm} for further details).

\section{The code capability: an example}
We remind the reader that the code is able to compute amplitudes up to NLO, as stated previously. To show the capability of the code, we will get all the possible contributions to the one loop diagrams including those with $\lagd^{(2)}$ and $\lagd^{(4)}$ vertices. 

\begin{figure}[]
\begin{center}
\includegraphics[width=0.17\textwidth]{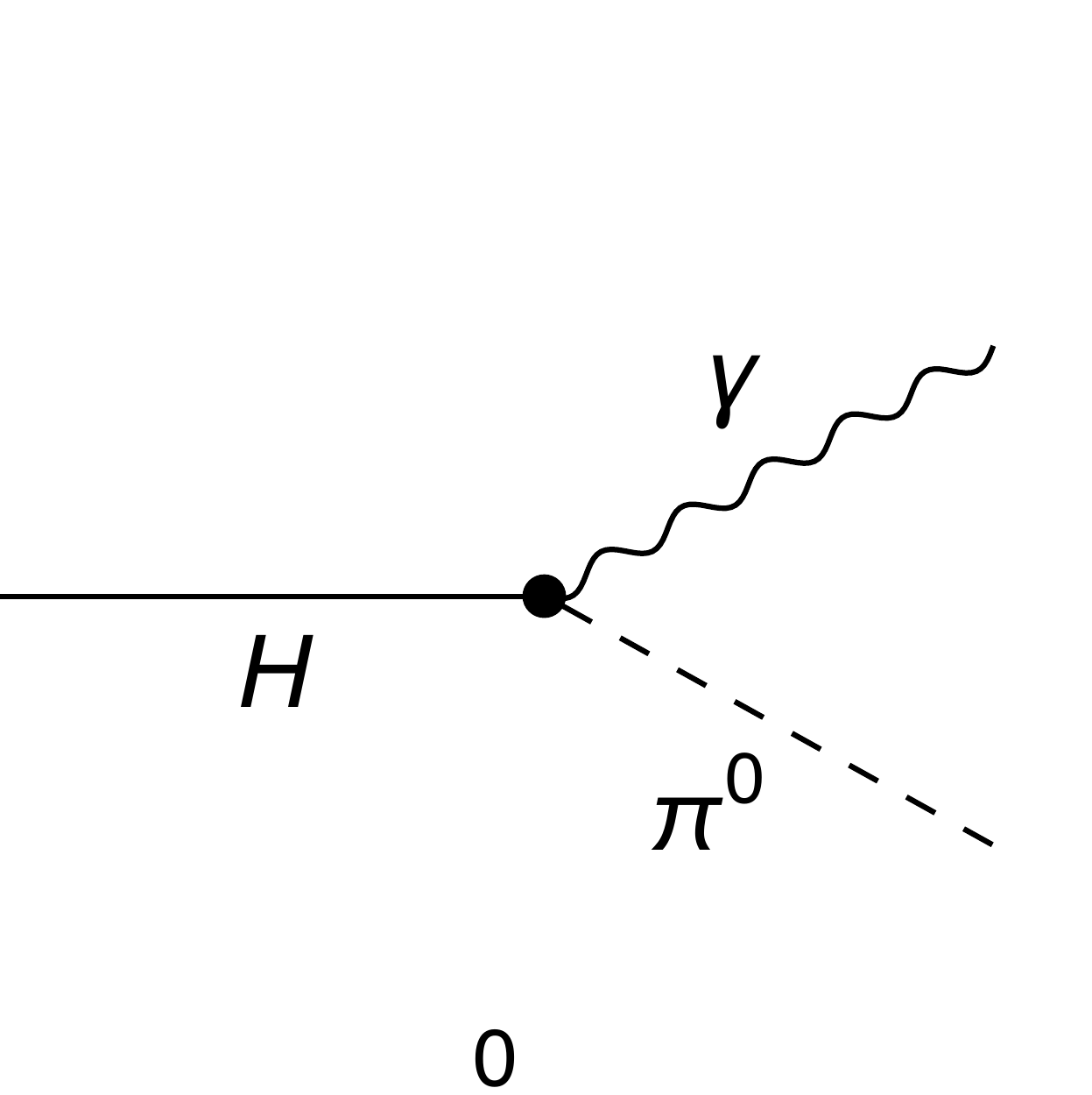}
\end{center}
\caption{\small $h(p_1)\; \longrightarrow \; \gamma(k_1,\epsilon_1)\; \pi_0(k_2)$} tree-level diagram $\mathcal{M}_0$.
\label{fig: H_APi0_0 tree}
\end{figure}

We have asked our code for the process ${h(p_1)\rightarrow A^\mu(k_1)\;\pi^0(k_2)}$. The tree level contribution is given by the diagram in Fig.~\ref{fig: H_APi0_0 tree}, with amplitude, 
\begin{equation}
\hspace*{-0.5cm}    \mathcal{M}_{0}=\frac{{e} \, \tilde{\mathcal{F}}_{3,0} }{v^2}\left[(s-m_h^2)(p_1\varepsilon^*_{k_1})+(t-m_h^2)(k_2\varepsilon^*_{k_1})\right] \, , 
\end{equation}

\begin{figure}[]
\begin{center}
\includegraphics[width=0.17\textwidth]{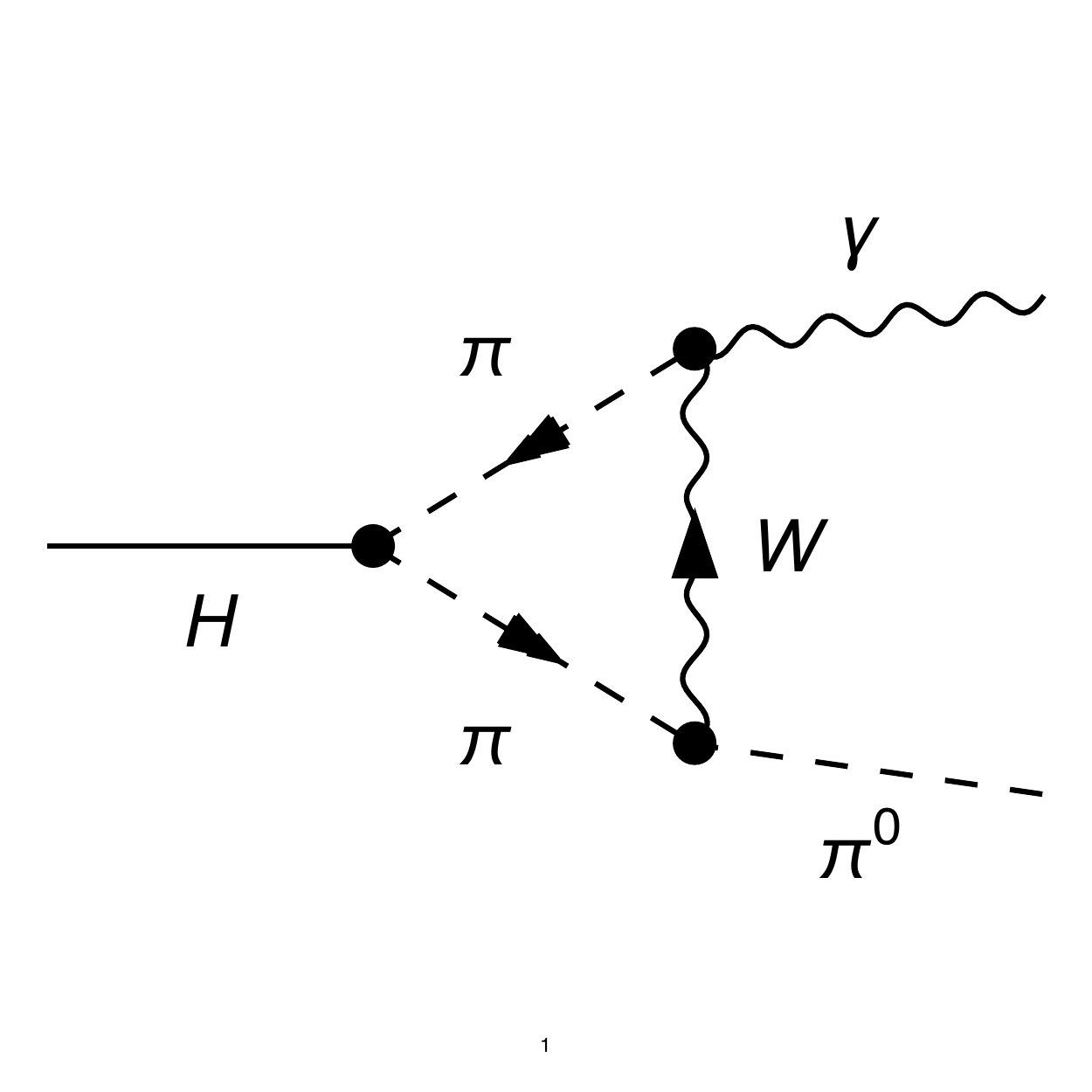}
\end{center}
\caption{\small $h(p_1)\; \longrightarrow \; \gamma(k_1,\epsilon_1)\; \pi_0(k_2)$} one-loop diagram $\mathcal{M}_1$.
\label{fig: H_APi0_L1}
\end{figure}

To go one step further, we have also asked for the one-loop contributions to the process. 
The code returns then 179 different one-loop diagrams and the corresponding amplitude for $\lagd=\lagd^{(2)}+\lagd^{(4)}$. This result is absurdly huge and there will be no point in presenting it in this paper, but the reader has it available in the ``EWET FeynCalc.nb'' notebook in~\cite{tfm}.  
At $\mathcal{O}(p^4)$, nonetheless, there are only one-loop diagrams with $\lagd^{(2)}$ vertices and one has just 22 diagrams.
For illustration, one of them is given in Fig.~\ref{fig: H_APi0_L1}, with the amplitude,
\begin{multline}
   \mathcal{M}_{1}=-\int\frac{d^dq}{(2\pi)^d}\frac{a\, e^3 \left(k_1 q+k_2 q-q^{2}\right) }{16 \pi^4
   {s_W}^2 q^{2} \left(-2
   k_1 q+k_1^{2}+q^{2}-m_W^2\right)}\cdot\\
   \cdot\frac{\left(k_1\varepsilon^*_{k_1}+2k_2 \varepsilon^*_{k_1}-q \varepsilon^*_{k_1}\right)}{ \left(2k_1 k_2-2k_1 q+k_1^{2}-2 k_2 q+k_2^{2}+q^{2}\right)}\, .
\end{multline}

\section{Conclusion}
With this work we have been able to create an automatized code \cite{tfm} capable of calculating any term of the Lagrangians $\lagd^{(2)}$ and $\lagd^{(4)}$. This allows us to calculate amplitudes with all the contributions up to $\mathcal{O}{(p^4)}$. We have build the part of the theory containing only the Higgs, the Goldstones and the EW bosons. The next step would be including the fermions, the ghosts and the gluons. 
This should be rather simple, as we will use the formalism already implemented in this version~1.

The goal of the code is to ease and quicken calculations in the framework of the EWChL. 
With the \textit{.mod}, \textit{.gen} and \textit{.pars} files one can start working directly with the vertices, saving the time of extracting them from the Lagrangian. These files can be read by \textsc{FeynCalc} to get amplitudes, integrate loops, export as UFO for MG5, etc.

We hope this code can help the high-energy physics community when further studying the Higgs Effective Theory. 
We expect to provide a more complete version with the extra sectors of the Lagrangian: fermions, gluons, ghosts and even resonances. 
Finally, as by-product, we intend to provide the analogous code for Chiral Perturbation Theory in QCD.

\end{document}